\documentclass{jfm}

\usepackage{graphicx}
\usepackage{newtxtext}
\usepackage{newtxmath}
\usepackage{natbib}
\usepackage{hyperref}
\hypersetup{
    colorlinks = true,
    urlcolor   = blue,
    citecolor  = black,
}

\newcommand{\RomanNumeralCaps}[1]

\title{Visibility network analysis of large-scale intermittency in convective surface layer turbulence}

\author{Subharthi Chowdhuri\aff{1}
  \corresp{\email{subharthi.cat@tropmet.res.in}},
  Giovanni Iacobello\aff{2,3}
 \and Tirtha Banerjee\aff{4}}

\affiliation{\aff{1}Indian Institute of Tropical Meteorology, Ministry of Earth Sciences, Pune, 411008, India
\aff{2}Department of Mechanical and Aerospace Engineering, Politecnico di Torino, 10129, Torino, Italy
\aff{3}Department of Mechanical and Materials Engineering, Queen’s University, Kingston, ON, K7L 3N6, Canada
\aff{4}Department of Civil and Environmental Engineering, University of California, Irvine, CA, 92697, USA}

\begin{document}
\maketitle

\begin{abstract}
Large-scale intermittency is a widely observed phenomenon in convective surface layer turbulence that induces non-Gaussian temperature statistics, while such signature is not observed for velocity signals. Although approaches based on probability density functions have been used so far, those are not able to explain to what extent the signals' temporal structure impacts the statistical characteristics of the velocity and temperature fluctuations. To tackle this issue, a visibility network analysis is carried out on a field-experimental dataset from a convective atmospheric surface layer flow. Through surrogate data and network-based measures, we demonstrate that the temperature intermittency is related to strong non-linear dependencies in the temperature signals. Conversely, a competition between linear and non-linear effects tends to inhibit the temperature-like intermittency behaviour in streamwise and vertical velocities. Based on present findings, new research avenues are likely to be opened up in studying large-scale intermittency in convective turbulence.
\end{abstract}

\section{Introduction}
\label{sec:intro}
Intermittency in turbulence is defined as a phenomenon generating fixed-point time series (e.g., of velocity or scalars) that are mostly quiescent, but occasionally there are occurrences of high peaks (large intensities) that last for a small duration \citep{batchelor1949nature,batchelor1953theory}. In general, two main types of intermittency appear in fully developed turbulent flows:  small-scale and large-scale intermittency \citep{tsinoberessence}. The small-scale intermittency is defined as the phenomenon where the statistics of the increments of turbulent quantities (e.g., velocity or other scalars) reflect a tendency, at scales comparable to the inertial and dissipation range, to burst to large values with a much greater occurrence than what would arise from a Gaussian process \citep{sreenivasan1997phenomenology,lohse2010small,shnapp2021small}. To assess small-scale intermittency, the scale dependence of the PDF or the moments of the velocity increments (structure functions of different orders) are typically evaluated. Moreover, other techniques such as multifractal analysis \citep{sreenivasan1991fractals,dupont2020experimental}, conditional wavelet analysis \citep{katul1994intermittency,matsushima2021wavelet}, or Hilbert-Huang transform \citep{huang2008amplitude} have been employed to study small-scale intermittency.\\
On the other hand, large-scale intermittency is defined as the phenomenon where the high-intensity fluctuations in the single-point PDFs of a turbulent flow occur with a significantly non-Gaussian probability \citep{majda1999simplified}. In single-point PDFs, the high-intensity fluctuations are typically associated with the large-scale motions (comparable to the integral or energy-containing scales). Hence, the broad non-Gaussian tails in such PDFs indicate highly-energetic activities occurring due to the occasional passage of the coherent structures over the measurement probe \citep{majda1999simplified,feraco2018vertical}.

The phenomenon of large-scale intermittency in convective flows attracted attention when the temperature measurements in Rayleigh-B\'{e}nard convection experiments exhibited non-Gaussian fluctuations, showing exponential PDF at sufficiently high Rayleigh numbers \citep{castaing1989scaling,siggia1994high,camussi2004temporal}. This finding was intriguing, since it was long expected that the single-point PDF of turbulent fluctuations would exhibit Gaussian distribution due to the central-limit theorem of statistics \citep{townsend1947measurement,batchelor1953theory,majda1999simplified}. On this note, \citet{jimenez1998turbulent} showed through a spectral model that the PDFs of velocity fluctuations in homogeneous turbulence do not necessarily have to be Gaussian, but can be slightly sub-Gaussian depending on the organization in the flow. Nevertheless, large-scale intermittency of temperature fluctuations has remained an enigmatic yet ubiquitous feature of Rayleigh-B\'{e}nard convection, where the continuing appearance of high-intensity fluctuations in the temperature field do not follow the Gaussian statistics \citep{roche2020ultimate}. Recently, \citet{he2018dynamic} and \citet{wang2019turbulent} theoretically showed that the non-Gaussianity in temperature for Rayleigh-B\'{e}nard convection is associated with thermal plumes, which are intermittently emitted from thermal boundary layers and carry temperature fluctuations towards different regions of the flow. Moreover, non-Gaussian temperature PDFs are also a remarkable feature of convective (i.e., buoyancy-dominated) atmospheric surface layer (ASL) flows \citep{chu1996probability,liu2011probability,lyu2018high}. The single-point PDFs of temperature fluctuations in convective ASL flows are more complex and display neither Gaussian nor exponential characteristics \citep{liu2011probability}. Besides, the non-Gaussianity in temperature fluctuations disappears as the ASL flow becomes shear-dominated where the temperature behaves like a passive scalar \citep{chu1996probability,lyu2018high}. 

Although the above statistical features of the temperature signals in a convective ASL flow have been widely observed, the origin of large-scale intermittency in such flows represents an issue that has received far less attention than small-scale intermittency, which instead has been the subject of several studies. Since large-scale intermittency has been mainly characterised by single-point PDFs so far \citep{majda1999simplified}, the role of the signal's temporal structure (i.e., the specific arrangement of data in time that results due to the passage of the coherent structures) on large-scale intermittency has remained uninvestigated. Accordingly, the connection between the organised coherent structures in ASL flow dynamics and large-scale intermittency has not been fully clarified yet. It is also not clear why the velocity signals in both buoyancy- and shear-dominated ASL flows appear to be near-Gaussian and do not display temperature-like intermittent behaviour. Particularly, the following research questions are largely unexplored, which have serious implications towards the turbulent flux modelling in convective ASL flows: (i) How does the temporal structure of the temperature and velocity signals evolve as the relative roles of shear and buoyancy changes in a convective ASL flow? (ii) What role does the flow organisation play in such evolution, especially to the large-scale intermittency as observed in the temperature signals? 

In order to evaluate impact of the temporal structure on large-scale intermittency, linear and non-linear effects represent two important features that need to be considered. In atmospheric turbulence, when a time series of velocity or any scalars is measured at a particular location, the temporal structure of the time series can be related to the effect of the organised flow patterns which pass over that location. When such patterns traverse the measurement location, they induce relationships between the points of the measured time series. To quantify these relationships, the simplest measure is the auto-correlation function (or alternatively the Fourier spectrum), which only describes the linear nature of the relationship. However, if linear functions cannot fully account for the dependencies present between the signal points, it can be concluded that the non-linear effects are present in the time series.\\
In our study, since a large Reynolds number ASL flow is considered whose governing equations are non-linear, it is apparent that the passage of the organised flow patterns will induce both linear and non-linear dependencies, affecting the temporal structure of the velocity and temperature fluctuations. Such non-linearities cannot be explained by only investigating the auto-correlation or Fourier amplitude spectrum, and therefore requires advanced techniques. Moreover, an important question arises, i.e. whether the effect of such non-linear dependencies will be similar for both the velocity and temperature signals. One may trivially assume the effect will be no different since the velocity and temperature signals are part of the same system governed by the non-linear equations. Nevertheless, this assumption needs to be checked, and the quantification of the effect of non-linearity requires a metric sensitive to the non-linear structure of the time series.
 
\begin{figure}
    \centering
    \includegraphics{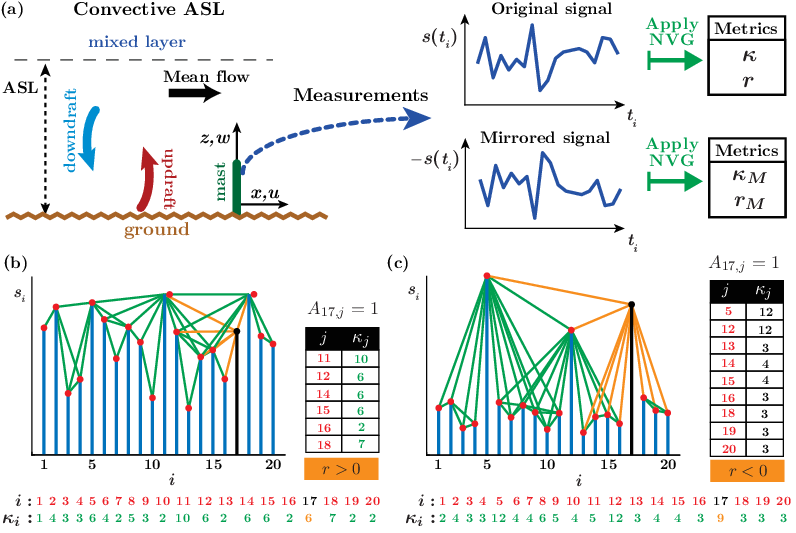}
    \caption{(a) Schematic of a convective ASL flow setup (left), and the NVG construction (right) from original and mirrored signals, $s_i$ and $-s_i$, respectively. (b) and (c) show two examples of signals, $s_i$, depicted as vertical blue bars, and the corresponding NVGs: (b) non-intermittent signal; (c) highly-intermittent signal. Nodes and links of the networks are depicted as red filled circles and green straight lines, respectively. The degree values, $\kappa_i$, for each node $i=1,\dots,20$ are listed in the bottom. In both panels, node $i=17$ is highlighted in black and its links in orange. The index, $j$, of neighbours of $i=17$ and the corresponding degree, $\kappa_j$, are also listed in two tables.}
    \label{fig:methods}
\end{figure}  

To address this issue, a network-based approach relying on natural visibility graph (NVG) is employed on a field-experimental dataset in a convective ASL flow (see figure \ref{fig:methods}a for a schematic). The NVG is used since it maps a time series into a network whose features (e.g., network metrics) are sensitive to the temporal structure of the mapped signal \citep{lacasa2008time} and therefore to a specific flow organisation. The NVG has been used for studying different vortical and turbulent flows \citep{iacobello2020review}, including wall-turbulence \citep{liu2010statistical, iacobello2018visibility,iacobello2021large}, jets and plumes \citep{charakopoulos2014application, iacobello2019experimental}, and turbulent combustors \citep{murugesan2015combustion, gotoda2017characterization}. The details of this method and the dataset are described in \S\,\ref{sec:methods}. The results of the analysis are presented in \S\,\ref{sec:results}, where we show through various network metrics that the velocity and temperature signals are characterised by different temporal structures. Finally, in \S\,\ref{sec:conclusions} conclusions are drawn and future research directions are provided.

\section{Methodology and dataset}
\label{sec:methods}
    \subsection{Visibility graph and network metrics}
    \label{visibility}
In order to capture the changes in the temporal structure of the signals, the NVG approach is employed, where network nodes correspond to temporal instants, $i=1,\dots,N$, of a time series, $s(t_i)$, while links are active if a convexity criterion is satisfied \citep{lacasa2008time}. Specifically, two nodes $\lbrace i,j\rbrace$, corresponding to signal values $\lbrace t_i, s(t_i)\rbrace$ and $\lbrace t_j, s(t_j)\rbrace$, are linked with each other if the condition (\ref{eq:visibility}) is fulfilled for any $t_i<t_n<t_j$:

\begin{equation}\label{eq:visibility}
	s(t_n)<s(t_j)+\left(s(t_i)-s(t_j)\right)\frac{t_j-t_n}{t_j-t_i}.
	\end{equation}

For instance, figure \ref{fig:methods}b-c show two synthetic signals, $s(t_i)$, and their corresponding visibility networks (depicted as red nodes and green links). The information on the link activation among nodes is then stored in the (binary) adjacency matrix, whose entries $A_{i,j}$ are equal to $1$ if there is a link between nodes $\lbrace i,j\rbrace$ with $i\neq j$, and $0$ otherwise \citep{newman2018networks}. Although the NVG algorithm is typically used as a convexity criterion, it can also be employed as a \textit{concavity} criterion applying (\ref{eq:visibility}) to the mirrored signal $-s(t)$ (figure \ref{fig:methods}a).   

Visibility graphs represent a class of network-based tools that, owing to its parameter-free nature and simplicity of implementation, has been largely adopted so far for signal analysis \citep{zou2019complex}. In particular, visibility graphs have shown to be able to inherit the structure of the signal, also accounting for non-linear effects \citep{lacasa2008time, manshour2015fully, manshour2015complex, iacobello2021large}. However, as per any other tool for time series analysis, some limiting aspects should be considered when visibility graphs are employed, such as the choice of the sampling frequency or the series length \citep{donner2012visibility}. Specifically, sampling frequency should be sufficiently high to capture the smallest scales of interest, while the signal length should be sufficiently long to capture the largest (temporal) scales. Recently, \citet{iacobello2021large} also evaluated the effect of high-frequency noise on experimental data on the visibility graph analysis of large-to-small scale frequency modulation in wall-bounded turbulence, showing a good robustness of the method. Finally, we note that the NVGs are insensitive to affine transformations of the signal, namely to shifting and re-scaling operation on the series axes \citep{lacasa2008time}, as a consequence of the linearity of the right-hand-side of inequality (\ref{eq:visibility}).

In order to characterise a network structure, and in turn the mapped signal, a set of network metrics have to be selected. One of the simplest metrics is the degree centrality, $\kappa_i$, defined as the number of neighbours of $i$, i.e., the nodes linked to $i$ \citep{newman2018networks}. As shown in figure \ref{fig:methods}b-c, the presence of locally convex intervals leads to the appearance of high-degree (i.e., highly-connected) nodes, which are typically associated with peaks in the signal. A point of a time series is here said to be a peak if it is a local (or global) maximum of the signal \citep{iacobello2018visibility}, namely its magnitude is comparable with the maximum excursion of the series (i.e., the difference between the maximum and the minimum values). A pit is the counterpart of a peak, namely a local (or global) minimum of the signal. Therefore, as network metrics (e.g., the degree centrality) are employed to characterise the signals' temporal structure, the ratio of the metrics for both NVG variants (i.e., built on the original and mirrored signals) can be used to study peaks-pits asymmetry in the signals \citep{hasson2018combinatorial}. In particular, the presence of peak-pit asymmetry in quantities (e.g., velocity and temperature) characterising a flow field, physically represents the tendency of the flow to display strong positive or negative fluctuations \citep{iacobello2019experimental}.

With the aim to quantify large-scale intermittency, the assortativity coefficient, $r$, is employed. $r$ is defined as the Pearson correlation coefficient between the degree of the nodes at the ends of each link, namely $r= \textrm{cov}\left[\kappa_i,\kappa_j\right]/ \left(\sigma\left[\kappa_i\right] \sigma\left[\kappa_j\right]\right)$, $\forall i,j$ so that $A_{i,j}=1$ \citep{newman2018networks}. Here, $\textrm{cov}[\bullet,\bullet]$ and $\sigma[\bullet]$ are the covariance and standard deviation, respectively. For example, in figure \ref{fig:methods}b-c, the indices, $j$, of the neighbours of representative node $i=17$ are listed together with the corresponding degree values, $\kappa_j$. In figure \ref{fig:methods}b, $\kappa_j$ values are similar to $\kappa_i=6$, while in figure \ref{fig:methods}c $\kappa_j$ values are dissimilar to $\kappa_i=9$. In the former case, nodes tend to link with other similar nodes so the $r$ value tends to increase (toward positive values); in the latter case, instead, nodes tend to link with other dissimilar nodes so the $r$ value tends to decrease (towards negative values).\\
\citet{iacobello2019experimental} showed that $r$ is an effective metric in quantifying intermittency in experimental signals of a meandering passive-scalar plume. Figure \ref{fig:methods}c shows an example of temporal structure of an intermittent signal (conceptual diagram). This structure can be compared with a not-so intermittent signal (figure \ref{fig:methods}b), where the subsequent values of the signal are more closer to each other. In figure \ref{fig:methods}c, peaks correspond to network hubs (high $\kappa$), which are mostly linked to nodes with lower degree values. Therefore, for intermittent signals, the corresponding NVGs tend to show smaller $r$ values tending towards negative values. In general, lower $r$ values correspond to strongest intermittent behaviours in the signal \citep{iacobello2019experimental}. Contrarily, low degrees of intermittency can be linked to higher and often positive values of $r$, as in figure \ref{fig:methods}b. Accordingly, the asymmetry between the intermittent behaviour of peaks and pits can be captured by evaluating the assortativity values of visibility graphs built from the original signals, $r$, and mirrored signals, $r_M$.    

\subsection{Dataset description}
\label{dataset}
We applied NVG on the dataset from the Surface Layer Turbulence and Environmental Science Test (SLTEST) experiment, which was conducted over a homogeneous and flat terrain at the Great Salt Lake desert in Utah, USA (40.14$^\circ$ N, 113.5$^\circ$ W). The experiment ran continuously for nine days from 26 May 2005 to 03 June 2005. In this experiment, nine time-synchronised sonic anemometers (CSAT3, Campbell Scientific, Logan, USA) were mounted on a 30-m mast at heights $z=\lbrace$1.4, 2.1, 3, 4.3, 6.1, 8.7, 12.5, 17.9, 25.7$\rbrace$ m, levelled to within $\pm$ 0.5$^\circ$ from the true vertical. Note that the lowest measurement height ($z=$ 1.4 m) at the SLTEST site was more than three orders of magnitude larger than the aerodynamic roughness length, which was $z_{0}\approx$ 5 mm \citep{metzger2007near}. Each of these nine anemometers measured the three wind components and sonic temperature at a sampling frequency of 20 Hz. Other details of the set-up and the instruments can be found in previous works \citep[e.g.,][]{hutchins2007evidence,chowdhuri2019revisiting}. 

In a convective ASL flow, the standard practice is to compute the turbulent statistics over a 30-min period \citep{panosfsky1984atmospheric,kaimal1994atmospheric}. Accordingly, the data from all the nine sonic anemometers were divided into 30-min segments, with each segment containing the 20-Hz measurements of the three wind components and the sonic temperature. Thereafter, we rotated the coordinate systems of all the nine sonic anemometers in the streamwise direction by applying the double-rotation method of \citet{kaimal1994atmospheric} for each 30-min period. The turbulent fluctuations in the wind components ($u^{\prime}$, $v^{\prime}$, and $w^{\prime}$ in the streamwise, cross-stream, and vertical directions respectively), and in the sonic temperature ($T^{\prime}$) were calculated after removing the 30-min linear trend from the associated variables. To select the 30-min periods for the NVG analysis, we focused only on rain-free daytime convective periods and followed the detailed data selection methods as outlined in \citet{chowdhuri2019revisiting}. 

In ASL flows, when turbulence is primarily dominated by buoyancy or shear, the time-averaged momentum or heat flux values become small (i.e., tend to zero). Although the vertical profiles of the heat and momentum fluxes in such flows should be constant with height \citep{panosfsky1984atmospheric}, due to the smallness of the flux values, vertical variations in the flux profiles are often noticed \citep{stiperski2018dependence, chowdhuri2019empirical}. Therefore, to include enough data that span from buoyancy-dominated to shear-dominated turbulence, we selected 30-min periods with at most a 40\% vertical variation in the heat and momentum fluxes with respect to the corresponding flux values at the lowest $z$. As a result, for all the selected 30-min periods from the convective conditions, a total of 1224 time series were collected in the range $10^{-3}<-\zeta<10^2$, where $-\zeta=z/L$ is the stability parameter, $z$ is the measurement height, and $L$ is the Obukhov length. Since $L$ is the height where the turbulence production due to buoyancy equals the production due to shear, $-\zeta>1$ denotes ASL flows that are largely dominated by buoyancy (highly-convective conditions). On the other hand, $-\zeta<0.1$ values conventionally indicate that the ASL flows are mostly dominated by shear (near-neutral conditions). 

\section{Results}
\label{sec:results}
This section reports the results of visibility networks related to the streamwise and vertical velocity fluctuations ($u^{\prime}$ and $w^{\prime}$) and temperature fluctuations ($T^{\prime}$) in a convective ASL flow. Note that here $\bullet^\prime$ notation indicates temporal fluctuations, i.e., the variables after the 30-min linear trend is removed. These three variables are chosen since they constitute the turbulent momentum and heat fluxes ($u^{\prime}w^{\prime}$ and $w^{\prime}T^{\prime}$) and hence are important quantities in the ASL turbulence dynamics.

\subsection{Intermittency via network assortativity}
\label{subsec:res_assort}
\begin{figure}
	 	\centerline{\includegraphics{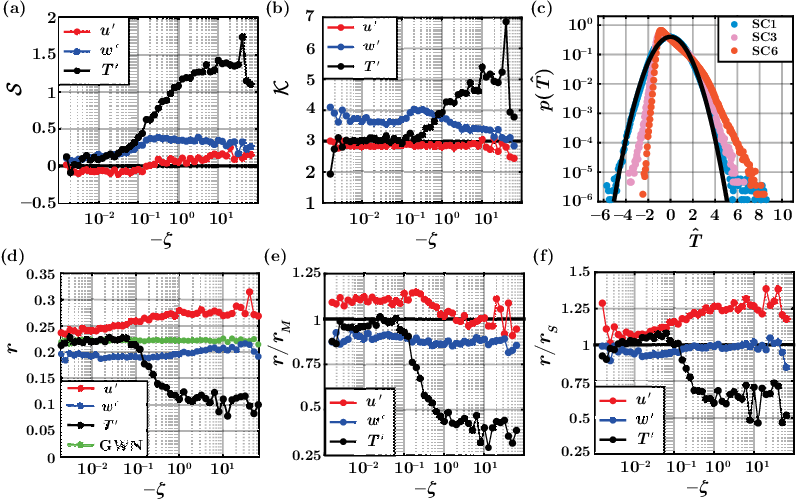}}
	  \caption{(a) Skewness, $\mathcal{S}$, and (b) kurtosis, $\mathcal{K}$, of $u^{\prime}$, $w^{\prime}$, and $T^{\prime}$ against the stability parameter, $-\zeta$. In (c) the PDF, $p(\hat{T})$, of the normalised temperature fluctuations, $\hat{T}=T^{\prime}/\sigma_{T}$ (where $\sigma_{T}$ is the standard deviation of $T^{\prime}$), are shown for the three stability classes (SC): SC1, $0<-\zeta<0.1$; SC3, $0.3<-\zeta<0.5$; SC6, $-\zeta>2$. The solid black line corresponds to the standard Gaussian distribution. (d) Assortativity coefficient, $r$, from the $u^{\prime}$, $w^{\prime}$, and $T^{\prime}$ signals against $-\zeta$. The green line indicates the $r$ values for a Gaussian white noise (GWN). In (e) and (f) the ratios between $r$ from the original $u^{\prime}$, $w^{\prime}$, and $T^{\prime}$ signals and their mirrored ($r_{M}$) and shuffled ($r_{S}$) counterparts, respectively, are shown.}  
		\label{fig:assortat}
	\end{figure}

In order to characterise the statistical features of large-scale intermittency, we begin by investigating the PDF-based higher-order moments of the analysed signals. Figure \ref{fig:assortat}a-b show the skewness ($\mathcal{S}$) and kurtosis ($\mathcal{K}$) of the $u^{\prime}$, $w^{\prime}$, and $T^{\prime}$ signals against the stability ratio $-\zeta$. For the $T^{\prime}$ signals, $\mathcal{S}$ and $\mathcal{K}$ significantly exceed their Gaussian values (0 and 3, respectively) for the highly-convective conditions ($-\zeta>1$). Whereas, in the near-neutral regime ($-\zeta<0.1$), the $T^{\prime}$ signals remain almost Gaussian. On the other hand, for the $u^{\prime}$ signals the $\mathcal{S}$ and $\mathcal{K}$ values are very close to their Gaussian counterparts irrespective of stability. However, for the $w^{\prime}$ signals a slight deviation from Gaussianity is observed as compared to the $u^{\prime}$ signals. 

The near-Gaussian behaviour of the $u^{\prime}$ and $w^{\prime}$ signals suggest that the velocity fluctuations are characterised by a low degree of large-scale intermittency, which agrees with previous studies \citep[e.g.,][]{chu1996probability}. But, in the buoyancy-dominated regime ($-\zeta>1$), the large skewness and kurtosis of the temperature signals inform that there are intermittent bursts of large positive values amidst the frequent occurrences of small negative fluctuations. To confirm this further, in figure \ref{fig:assortat}c we show the temperature PDFs for the three representative stability classes. We notice that at $-\zeta>2$, the $T^{\prime}$ PDF significantly differs from a Gaussian distribution with anomalous occurrences of large positive $T^{\prime}$ values.  

Despite the fact that PDF-based analyses of large-scale intermittency have been widely used, they cannot fully explain the role of flow organisation on the signal's intermittent characteristics \citep{tsinoberessence}. Contrarily, the assortativity coefficient, $r$, quantifies large-scale intermittency by retaining information on the temporal structure of the signals. In figure \ref{fig:assortat}d, the behaviour of $r$ is shown for the $u^{\prime}$, $w^{\prime}$, and $T^{\prime}$ signals against $-\zeta$. For $T^{\prime}$, the $r$ values undergo a gradual transition between two distinct plateaus of constant values for the near-neutral ($-\zeta<0.1$) and highly-convective ($-\zeta>1$) conditions. Particularly, in the near-neutral regime the $r$ values of $T^{\prime}$ remain close to a Gaussian white noise (GWN). Since a decrease in $r$ is sensitive to an increasing large-scale intermittency of the signal, the temperature fluctuations in a buoyancy-dominated ASL flow ($-\zeta>1$) appear more intermittent as compared to a shear-dominated ASL flow ($-\zeta<0.1$). We also observe that $r$ remains almost invariant with $-\zeta$ for $w^{\prime}$, while a slight increase in $r$ is observed for $u^{\prime}$.

The similarity between $r$ and PDF-based statistics as a function of $-\zeta$ validates our metric, which is able to capture the main features of the underlying dynamics. On this note, the NVGs can be further used to assess whether the large-scale intermittency in the signal is related to the positive or negative fluctuations via studying the assortativity coefficients of the original ($r$) and mirrored ($r_{M}$) signals. Figure \ref{fig:assortat}e shows the ratio $r/r_{M}$ that characterises the asymmetry in large-scale intermittency between positive and negative fluctuations from a network perspective. For $T^{\prime}$ signals, $r/r_{M}$ decreases from the shear-dominated regime towards the buoyancy-dominated regime. Physically this indicates that, for highly-convective regimes ($-\zeta>1$), the mirrored signal is less intermittent, given that $r_{M}>r$ (see Fig. S1 in the supplementary material). Moreover, since the changes in $r_{M}$ with $-\zeta$ remain significantly less as compared to $r$ (see Fig. S1), the behaviour of $r⁄r_{M}$ tends to mimic the behaviour of $r$. Therefore, in highly-convective conditions the large-scale intermittency in $T^{\prime}$ is dictated by positive fluctuations (i.e., peaks, $T^{\prime}>0$) rather than the negative ones (i.e., pits, $T^{\prime}<0$). This finding is consistent with the fact that the PDFs of $T^{\prime}$ in the highly-convective conditions show a heavy tail towards the positive fluctuations, as evident from figure \ref{fig:assortat}c. Moreover, the ratio $r/r_{M}$ for the two velocity components remains almost constant and approximately equal to 1, suggesting a peak-pit symmetry of the signals for any $-\zeta$ in agreement with results shown in figure \ref{fig:assortat}a-b. It is worth remarking that, despite the aforementioned overall agreement, the NVGs are able to isolate the intermittent behaviour of positive and negative fluctuations, while higher-order moments do not distinguish extreme events as peaks and pits. For instance, the kurtosis (or any other even-order moment) does not distinguish between the positive and negative extreme events, while the skewness does not account for only extreme events but both small and large values in the signal.

To further explore the role of the temporal structure of the signals on large-scale intermittency, we computed the ratios of $r$ values between the original and randomly shuffled signals, $r/r_{S}$. To compute $r_{S}$, one shuffled surrogate series was generated for each time series of $u^{\prime}$, $w^{\prime}$ and $T^{\prime}$; however, a larger number of surrogates do not lead to significant changes in $r_{S}$. Since random shuffling destroys the temporal dependencies while preserving the signal PDF (and therefore all the PDF-moments), the deviation from 1 in $r/r_{S}$ is an indication of how strong is the impact of the temporal structure on large-scale intermittency. Figure \ref{fig:assortat}f illustrates that $r/r_{S}$ significantly decreases for $T^{\prime}$ and increases for $u^{\prime}$ towards $-\zeta>1$, while no discernible change is observed for $w^{\prime}$. It is thus apparent that, in buoyancy-dominated regime, the flow organisation plays an important role to determine the temporal arrangement of $u^{\prime}$ and $T^{\prime}$ values in the signals. However, since shuffling destroys any dependencies in a time series, a more detailed analysis is required to shed light on the relative contribution of linear and non-linear effects undergoing in the flow, which impact the signal's intermittent features. 

\subsection{Role of non-linearity on convective dynamics}
\label{subsec:res_nonlin}
To investigate the role of non-linearity, first we investigate the behaviour of the degree distribution (i.e. the probability to find a node with a given degree value, $\kappa$), since it is strongly related to the temporal structure of the mapped signals \citep{lacasa2008time,liu2010statistical}.  In fact, the degree distribution from the NVG is sensitive to the signal's PDF and on the linear and non-linear dependencies in the signals \citep{manshour2015complex,iacobello2018visibility}. To account for the linear and non-linear effects, we evaluated the degree distributions of $u^{\prime}$, $w^{\prime}$, and $T^{\prime}$ from two different types of surrogate signals: (i) random shuffling, and (ii) iteratively adjusted amplitude Fourier transform (IAAFT). In IAAFT method, surrogate data are generated that do not contain non-linear effects but preserve the linear effects described by the auto-correlation in the time series \citep{lancaster2018surrogate}. To achieve such objective, the Fourier amplitudes of the time series are kept intact (thereby preserving the auto-correlation), but the associated Fourier phases are replaced by a random uniform distribution between 0 to 2$\pi$. The randomness in the Fourier phases destroys any non-linear structure of the time series. However, due to the randomisation of the Fourier phases the PDF of the time series is altered to be Gaussian. Hence, to preserve both PDF and amplitude spectrum, the Fourier amplitudes and the signal's PDFs are adjusted iteratively at each stage of phase randomisation until the resultant signal has the same power spectrum and the PDF as the original one. \citet{lancaster2018surrogate} provides a step-by-step implementation of the IAAFT algorithm, which has been also applied to turbulence studies \citep{poggi2004interaction,basu2007estimating,cabrit2013fundamental,keylock2017multifractal}. 

\begin{figure}
	 	    \centerline{\includegraphics{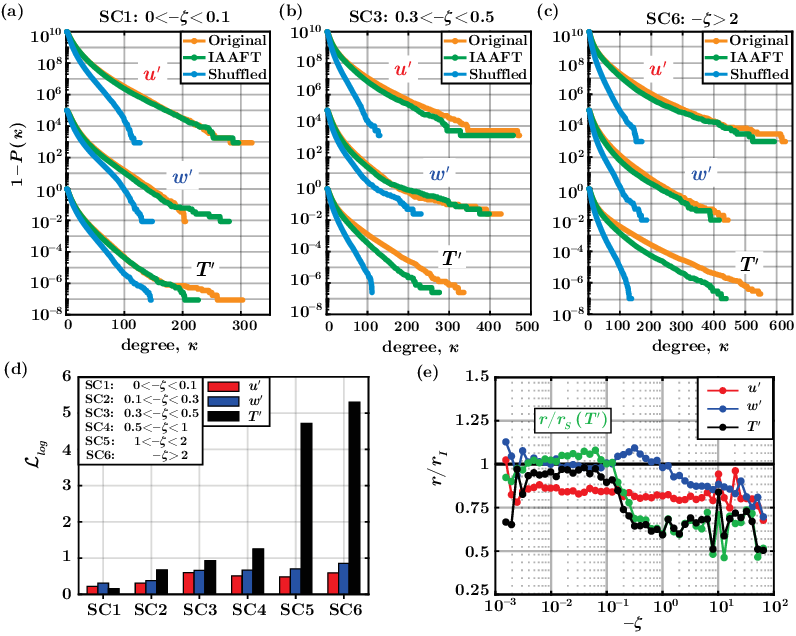}}
		    \caption{The complementary CDFs, $1-P(\kappa)$, of the degree, $\kappa$, from the original, shuffled, and IAAFT signals corresponding to three representative stability classes: (a) $0<-\zeta<0.1$, (b) $0.3<-\zeta<0.5$, and (c) $-\zeta>2$. For clarity purposes, the distributions are shifted vertically by five and ten decades for $w^{\prime}$ and $u^{\prime}$, respectively. (d) The L1-norm, $\mathcal{L}_{log}$, between the difference of original and IAAFT signals for the six stability classes (SC). (e) Ratio, $r/r_{I}$, against $-\zeta$ between $r$ from the original signals and $r_{I}$ from IAAFT signals. For comparison, the ratio $r/r_{S}$ of the $T^{\prime}$ signal (figure \ref{fig:assortat}f) is included in (e) as a green line.}  
		    \label{fig:deg_iafft}
	    \end{figure}

In figure \ref{fig:deg_iafft}a-c, we show the complementary cumulative distribution functions (cCDF) of the degree, $1-P(\kappa)$ (where $P(\kappa)$ is the degree CDF), from the original, shuffled, and IAAFT surrogates of $u^{\prime}$, $w^{\prime}$, and $T^{\prime}$, for three representative stability classes (see Appendix \ref{appA} for cCDFs of $\kappa$ from all the six stability classes). cCDFs are typically used to overcome the intrinsically fat-tailed behaviour of degree distributions and to highlight the contribution from high $\kappa$ values \citep{newman2018networks}. Irrespective of the stability class, the cCDFs of original signals (orange lines) and their randomly shuffled surrogates (blue lines) remain far apart from each other, revealing the strong effect of the signal's temporal structure on the network topology \citep{iacobello2018visibility}. On the other hand, the difference between the cCDFs from original and IAAFT signals (green lines in figure \ref{fig:deg_iafft}a-c) are less evident. Only for temperature, $T^{\prime}$, a remarkable difference in cCDFs is observed in the highly-convective regime (figure \ref{fig:deg_iafft}c), which systematically disappears as the near-neutral regime is approached (figure \ref{fig:deg_iafft}a).

The comparison between the original degree distributions and their IAAFT surrogates is useful to assess the net role of non-linearity on the signal's temporal structure, since these surrogates retain only the signal's PDF and its linear correlations. Therefore, the results of figure \ref{fig:deg_iafft}a-c suggest that the effects of non-linearity remain quite different on the temporal structure of velocity and temperature fluctuations, despite being part of the same non-linear system (i.e., atmospheric turbulence at a large Reynolds number). In order to quantify the net effect of non-linearity in $P(\kappa)$, we calculated the L1-norm of the difference between the logarithms of $P(\kappa)$, i.e., $\mathcal{L}_{\log}=||\log[P(\kappa)_{\rm{I}}]-\log[P(\kappa)_{\rm{O}}] ||_1$, where I and O subscripts refer to IAAFT and original signals, respectively. The L1-norm is a simple yet useful metric to compare the difference between two probability distributions \citep{hasson2018combinatorial}. The logarithm is taken since $P(\kappa)$ have an exponential tail so that in computing the L1-norm we balance the contribution to $P(\kappa)$ coming from low and high $\kappa$ values. 

In figure \ref{fig:deg_iafft}d we present the $\mathcal{L}_{\log}$ values between the original and IAAFT signals as a bar-plot for the same six stability classes. In the near-neutral regime ($0<-\zeta<0.1$) the impact of non-linearity is rather low and similar for all the signals. For the $T^{\prime}$ signals, the $\mathcal{L}_{\log}$ values gradually increase as the highly-convective regime ($-\zeta>2$) is approached. This implies that the temporal structure of the temperature fluctuations becomes highly-sensitive to the non-linear dependencies in such conditions. On the other hand, for the velocity signals, we observe that the L1-norm for the $w^{\prime}$ signals is slightly larger than the $u^{\prime}$ signals. This outcome suggests that the $w^{\prime}$ signals appear more affected than the $u^{\prime}$ signals by non-linear dependencies for large $-\zeta$ values.

So far, we have reported through the degree distribution that the NVGs not only are sensitive to the full temporal structure (figure \ref{fig:deg_iafft}a-c), but are also able to highlight the impact of non-linear dependencies in the signals (figure \ref{fig:deg_iafft}d). Therefore, since the NVG changes with non-linearity, the network assortativity coefficient, $r$, -- which accounts for signal intermittency -- is also expected to change. To evaluate such a variation in $r$, figure \ref{fig:deg_iafft}e illustrates the ratio, $r/r_{I}$, computed between $r$ for the original signals, and $r_I$ for its IAAFT surrogates. Note that, in accordance with \citet{poggi2004interaction}, we used 10 IAAFT surrogates for each signal to compute the $r_I$ values. Values of $r/r_{I}$ close to 1 indicate that there is a negligible impact of non-linearity on the signal intermittency, since $r$ accounts for both linear and nonlinear effects, while $r_{I}$ only for linear effects. For the $T^{\prime}$ signals, $r/r_{I}$ decreases from 1 in the shear-dominated regime to around 0.6 in the buoyancy-dominated regime. Interestingly, $r/r_I$ for the $T^{\prime}$ signals is almost coincident with the corresponding $r/r_S$ (green line in figure \ref{fig:deg_iafft}e), which accounts for both linear and non-linear effects on the signal's temporal structure. Hence, it indicates that the temperature intermittency -- especially in the highly-convective regime -- is mainly characterised by the non-linear dependencies in the signal, while the linear effects are negligible.

Contrarily, for the velocity signals the effect of non-linearity on $r$ is different. For $u^{\prime}$ signals, the ratio $r/r_{I}$ remains almost constant with $-\zeta$ (figure \ref{fig:deg_iafft}e), although its $r/r_{S}$ values increase with $-\zeta$ (figure \ref{fig:assortat}f). This suggests that the changes in the temporal structure of $u^{\prime}$ are primarily governed by linear dependencies -- described by the amplitude spectrum -- as the highly-convective regime is approached. For the ASL flows, it is known that there is an increase in large-scale spectral energy with respect to the small-scales, as the turbulence production due to buoyancy is increased \citep{wyngaard2010turbulence}. This relative increase in large-scale spectral energy results in smoother $u^{\prime}$ signals, given the small scales are less energetic. In general, smother signals correspond to more regular NVGs and higher assortativity values \citep{iacobello2018visibility, iacobello2019experimental,iacobello2021large}. Thus, the increasing trend shown in figure \ref{fig:assortat}f for $u^{\prime}$ is mainly related to linear effects and is a result of a change in the spectral energy at small and large scales. For $w^{\prime}$, on the other hand, we see that the ratio $r/r_I$ decreases in the highly-convective regime. Similar to $u^{\prime}$, as the convective regime is approached, the large-scale spectral energy increases with respect to the small-scales \citep{wyngaard2010turbulence}, thus making the $w^{\prime}$ signal smoother. Linear effects, therefore, would make the assortativity increase while (as shown in figure \ref{fig:deg_iafft}e) nonlinear effects make it decrease. The net effect is an apparent independence of the assortativity on the full temporal structure with $-\zeta$, as clearly observed in figure \ref{fig:assortat}f. In other words, there is a balance between the linear and non-linear effects which govern the temporal structure of the $w^{\prime}$ signals in the buoyancy-dominated regime.

\section{Discussion and conclusions}
\label{sec:conclusions}
The impact of the temporal structure -- including both linear and non-linear dependencies -- on large-scale intermittency in convective ASL flows is addressed for the first time through a visibility graph approach. The stability conditions have the strongest impact on the temporal structure of the $T^{\prime}$ signals, while their impact is weaker on the temporal structure of the $u^{\prime}$ and $w^{\prime}$ signals. Combining visibility graphs and surrogate data, we demonstrate that the intermittent structure of $T^{\prime}$ in convective conditions is associated with strong non-linear dependencies in the signal. In fact, in highly-convective ASL flows, the small negative fluctuations observed in $T^{\prime}$ are signature of the cold-downdrafts, which bring well-mixed air from aloft (the upper regions of the ASL flow or the mixed layer, see figure \ref{fig:methods}a) to the heights closer to the ground. On the other hand, the large positive temperature fluctuations can be related to warm-updrafts associated with the hot-plumes rising from the surface. Therefore, the observed intermittency in temperature could be due to a complex non-linear interaction between the cold-downdrafts and warm-updrafts, which results in a specific temporal pattern that appears to be intermittent. Put differently, in highly-convective ASL flows, a non-linear interplay between the temperature field and the underlying turbulent flow dynamics seems to be responsible for the large-scale intermittency in the temperature fluctuations.

As discussed in his seminal work, \citet{batchelor1953theory} observed that the non-linear terms in the governing equations of turbulence would influence the non-Gaussian behaviour of the velocity increments, i.e., the small-scale intermittency. Recently, \citet{zorzetto2018extremes} confirmed the non-linear character of the system is responsible for the small-scale intermittency in the statistics of temperature increments. However, our results illustrate that the non-linearity even affects the non-Gaussian characteristics of the single-point measurements of temperature fluctuations in highly-convective conditions (i.e., large-scale intermittency). This finding is not trivial because it implies that, for temperature fluctuations, the non-linear effects even persist at scales comparable to the energy-containing scales of motions that cause the highly-energetic activities in $T^{\prime}$, thus leading it to be non-Gaussian \citep{majda1999simplified}. Hence, it can be inferred that, since the non-linearity influences both the small- and large-scale intermittency in temperature fluctuations, these two phenomena seem to be coupled in the case of temperature signals.\\
Although it currently represents a conjecture, some experimental evidence exists that supports this picture. For instance, in Rayleigh-B\'{e}nard convection experiments, it has been found that the non-Gaussian statistics of temperature increments (statistics of temperature derivatives and dissipation rate) and temperature fluctuations are closely linked together \citep{belmonte1996thermal,emran2008fine}. Similarly, in a highly-convective ASL flow, the ramp-cliff structures (slow rises followed by a sudden drop) in the $T^{\prime}$ time series have been found to affect the small-scale and large-scale intermittency in such signals. In fact, while ramp-cliff structures are a signature of the presence of convective plumes inducing non-Gaussian temperature PDFs \citep{chu1996probability}, thus contributing to large-scale intermittency, they have also been shown to impact the small-scale intermittency of temperature during highly-convective conditions \citep{zorzetto2018extremes}.

Concerning velocity fluctuations, with the change in the stability ratio ($-\zeta$), the linear effects (related to the Fourier spectra) play a stronger role in determining the temporal structure of the $u^{\prime}$ and $w^{\prime}$ signals than $T^{\prime}$. Specifically, the change in the temporal structure of $u^\prime$ with stability is driven by a relative increase and decrease in the spectral energy at large and small scales, respectively. This leads $u^{\prime}$ signals to be less-intermittent than temperature signals, as revealed by higher assortativity values. On the other hand, a competition between the linear and non-linear effects inhibits temperature-like intermittency in the $w^{\prime}$ signals.

The results provided in this study represent a first attempt towards advancing the level of information of PDF-based analyses, thus fostering the possibility to improve the understanding of large-scale intermittency in convective ASL flows, a seldom-explored topic in the literature. More importantly, complex networks can be a powerful tool to assess the role of non-linearity on the organisation of coherent structures, and its impact on the turbulent fluxes. On this note, to further shed light on how the organised structures in a convective ASL flow affect the intermittent features of the heat and momentum flux signals, an explicit assessment of the inter-relations between the velocity and temperature signals would be required. This should also involve more advanced analyses, e.g., via multi-layer networks \citep{kivela2014multilayer}, whose application is out of the scope of the present manuscript and will be the subject of a future research. In conclusion, based on current findings and methodology, we do believe the present work could open up new research avenues in studying large-scale intermittency in convective flows.

\backsection[Acknowledgements]{SC is indebted to KG McNaughton for providing the SLTEST dataset.}

\backsection[Funding]{Indian Institute of Tropical Meteorology (IITM) is an autonomous institute fully funded by the Ministry of Earth Sciences, Government of India. TB acknowledges the funding support from the University of California Laboratory Fees Research Program funded by the UC Office of the President (UCOP), grant ID LFR-20-653572.Additional support was provided to TB by the new faculty start up grant provided by the Department of Civil and Environmental Engineering, and the Henry Samueli School of Engineering, University of California, Irvine.}

\backsection[Declaration of interests]{The authors report no conflict of interest.}

\backsection[Author ORCID]{S. Chowdhuri, 0000-0002-5518-7701; G. Iacobello, 0000-0002-0954-8545; T. Banerjee, 0000-0002-5153-9474}

\backsection[Author contributions]{SC and GI designed the study and carried out the analyses. GI prepared the figures and SC wrote the initial draft of the manuscript. GI and TB provided their corrections, comments, and suggestions. All the authors read the final draft and agreed to all the changes.}

\appendix
\section{Degree distributions for all stability classes}
\label{appA}
\begin{figure}
	 	    \centerline{\includegraphics{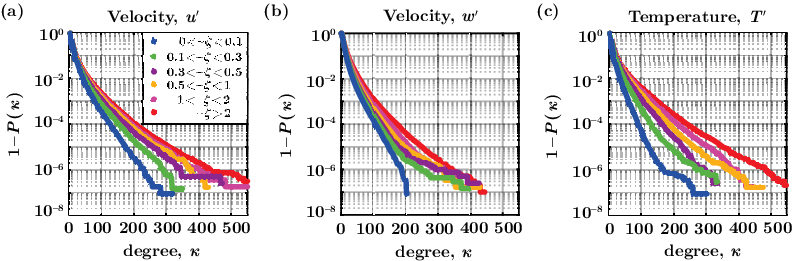}}
		    \caption{The complementary CDFs ($1-P(\kappa)$) of the degrees ($\kappa$) are shown for the (a) $u^{\prime}$, (b) $w^{\prime}$, and (c) $T^{\prime}$ signals corresponding to the six stability classes. The classifications of the stability classes according to the ranges in $-\zeta$ are in agreement with \citet{liu2011probability}. Each stability class contains over 100 number of 30-min runs, thus ensuring the statistical robustness of the degree distributions presented here.}  
		    \label{fig:deg_CDF}
	    \end{figure}

In this appendix, the behaviour of the cCDFs of the degree, $\kappa$, for six different stability classes is reported. This is to highlight the change in the overall temporal structure of the $u^{\prime}$, $w^{\prime}$, and $T^{\prime}$ signals as the ASL transitions from the buoyancy- to shear-dominated regime. The cCDFs are illustrated in figure \ref{fig:deg_CDF} in a log-linear plot for the three variables, showing exponential tails. For the $u^{\prime}$ and $T^{\prime}$ signals (figure \ref{fig:deg_CDF}a and c) the slopes of the distributions get gradually steeper as $-\zeta\rightarrow 0$, i.e. as the ASL approaches the near-neutral conditions. On the contrary, a small change is observed in the slopes of the $w^{\prime}$ signals (figure \ref{fig:deg_CDF}b). The results shown in figure \ref{fig:deg_CDF} are qualitatively in general accordance with the behaviour of $r$ as shown in figure \ref{fig:assortat}d  (which focuses on the intermittent structure of the signals). In fact, in both figures we observe stronger changes with $-\zeta$ for the $u^{\prime}$ and $T^{\prime}$ signals, and weaker variations in $w^{\prime}$.

\clearpage
\bibliographystyle{jfm}
\bibliography{ATL_Net.bib}
\end{document}